\documentclass[11pt]{article}
\usepackage{amssymb}
\usepackage{amsmath}
\usepackage[english]{babel}
\usepackage{graphicx}

\def\Journal#1#2#3#4{{#1} {\bf #2}, #3 (#4)}

\def\RNC{\em Rivista Nuovo Cimento}

\def\NIMA{{\em Nucl. Instrum. Methods} A}

\def\PLB{{\em Phys. Lett.}  B}
\def\PRL{\em Phys. Rev. Lett.}
\def\PRD{{\em Phys. Rev.} D}

\def\GaC{\em Gravitation and Cosmology}

\def\JETPL{\em JETP Lett.}

\def\CQG{\em Class. Quantum Grav.}
\def\APJ{\em Astrophys. J.}
\def\SCI{\em Science}
\def\MPLA{{\em Mod. Phys. Lett.}  A}
\def\IJTP{\em Int. J. Theor. Phys.}
\def\NJP{\em New J. of Phys.}
\def\JHEP{\em JHEP}
\def\EPHJ{\em Eur.Phys.J}
\def\BWP{\em Bled Workshops in Physics}

\def\s{{\,\rm s}}
\def\g{{\,\rm g}}
\def\eV{\,{\rm eV}}
\def\keV{\,{\rm keV}}
\def\MeV{\,{\rm MeV}}
\def\GeV{\,{\rm GeV}}
\def\TeV{\,{\rm TeV}}
\def\sv{\left<\sigma v\right>}
\def\({\left(}
\def\){\right)}
\def\cm{{\,\rm cm}}

\def\kpc{{\,\rm kpc}}
\def\beq{\begin{equation}}
\def\eeq{\end{equation}}
\def\bea{\begin{eqnarray}}
\def\eea{\end{eqnarray}}

\begin{document}

    \begin{center}
        \large \textbf{Dark Atoms of the Universe: towards OHe nuclear physics}
    \end{center}

    \begin{center}
   Maxim Yu. Khlopov$^{1,2,3}$, Andrey G. Mayorov $^{1}$, Evgeny Yu.
   Soldatov $^{1}$

    \emph{$^{1}$National Research Nuclear University "Moscow Engineering Physics Institute", 115409 Moscow, Russia \\
    $^{2}$ Centre for Cosmoparticle Physics "Cosmion" 115409 Moscow, Russia \\
$^{3}$ APC laboratory 10, rue Alice Domon et L\'eonie Duquet \\75205
Paris Cedex 13, France}

    \end{center}

\medskip

\begin{abstract}

The nonbaryonic dark matter of the Universe is assumed to consist of
new stable particles.
 A specific case is possible, when new stable particles bear ordinary
 electric charge and bind in heavy "atoms" by ordinary Coulomb
 interaction. Such possibility is severely restricted by
 the constraints on anomalous isotopes of light elements that
 form positively charged heavy species with ordinary electrons.
 The trouble is avoided, if stable particles $X^{--}$ with charge
 -2 are in excess over their antiparticles (with charge +2) and there are no stable
 particles  with charges +1 and -1. Then primordial helium, formed in Big Bang
 Nucleosynthesis, captures all $X^{--}$ in
 neutral "atoms" of O-helium (OHe).  Schrodinger equation for system of nucleus
 and OHe is considered and reduced to an equation of relative motion in a spherically symmetrical potential,
 formed by the Yukawa tail of nuclear scalar isoscalar attraction potential, acting on He beyond the nucleus,
 and dipole Coulomb repulsion between the nucleus and OHe at small distances between nuclear
 surfaces of He and nucleus. The values of coupling strength and mass of $\sigma$-meson,
 mediating scalar isoscalar nuclear potential, are rather uncertain. Within these
 uncertainties and in the approximation of rectangular potential wells and wall
 we find a range of these
 parameters, at which  the sodium nuclei have a few keV binding energy with OHe.
 The result also strongly depend on the precise value of parameter
 $d_o$ that determines the size of nuclei.
 At nuclear parameters, reproducing DAMA results, OHe-nucleus bound states can exist only for
 intermediate nuclei, thus excluding direct comparison with these results in detectors, containing
 very light (e.g. $^3He$) and heavy nuclei (like Xe).

\end{abstract}
\section{Introduction}
Ordinary matter around us consists of neutral atoms, in which
electrically charged nuclei are bound with electrons. Ordinary
matter is luminous because of electron transitions in atoms. It is
stable owing to stability of its constituents. Electron is the
lightest charged particle. It is stable due to conservation of
electromagnetic charge that reflects local gauge U(1) invariance.
Electromagnetic charge is the source of the corresponding U(1) gauge
field, electromagnetic field. Nuclei are stable because of stability
of nucleons. The lightest nucleon - proton - is the lightest baryon
and stable due to conservation of baryon charge. There is no gauge
field related with baryon charge. Therefore there are two examples
of stable charged particles of the ordinary matter: protected by
gauge symmetry and protected by conserved charge. This excursus in
known physics can give us some idea on possible constituents of dark
atoms, maintaining the dark matter of the Universe.

According to the modern cosmology, the dark matter, corresponding to
$25\%$ of the total cosmological density, is nonbaryonic and
consists of new stable particles. Such particles (see e.g.
\cite{book,Cosmoarcheology,Bled07} for review and reference) should
be stable, saturate the measured dark matter density and decouple
from plasma and radiation at least before the beginning of matter
dominated stage. The easiest way to satisfy these conditions is to
involve neutral elementary weakly interacting particles. However it
is not the only particle physics solution for the dark matter
problem and more evolved models of self-interacting dark matter are
possible. In particular, new stable particles may possess new U(1)
gauge charges and bind by Coulomb-like forces in composite dark
matter species. Such dark atoms would look nonluminous, since they
radiate invisible light of U(1) photons. Historically mirror matter
(see \cite{book,Okun} for review and references) seems to be the
first example of such a nonluminous atomic dark matter.

Glashow's tera-helium \cite{Glashow} has offered a new solution for
dark atoms of dark matter. Tera-U-quarks with electric charge +2/3
formed stable (UUU) +2 charged "clusters" that formed with two -1
charged tera-electrons E neutral [(UUU)EE] tera-helium "atoms" that
behaved like Weakly Interacting Massive Particles (WIMPs). The main
problem for this solution was to suppress the abundance of
positively charged species bound with ordinary electrons, which
behave as anomalous isotopes of hydrogen or helium. This problem
turned to be unresolvable \cite{Fargion:2005xz}, since the model
\cite{Glashow} predicted stable tera-electrons $E^-$ with charge -1.
As soon as primordial helium is formed in the Standard Big Bang
Nucleosynthesis (SBBN) it captures all the free $E^-$ in positively
charged $(He E)^+$ ion, preventing any further suppression of
positively charged species. Therefore, in order to avoid anomalous
isotopes overproduction, stable particles with charge -1 (and
corresponding antiparticles) should be absent, so that stable
negatively charged particles should have charge -2 only.

Elementary particle frames for heavy stable -2 charged species are
provided by: (a) stable "antibaryons" $\bar U \bar U \bar U$ formed
by anti-$U$ quark of fourth generation \cite{Q,I,lom,Khlopov:2006dk}
(b) AC-leptons \cite{Khlopov:2006dk,5,FKS}, predicted in the
extension \cite{5} of standard model, based on the approach of
almost-commutative geometry \cite{bookAC}.  (c) Technileptons and
anti-technibaryons \cite{KK} in the framework of walking technicolor
models (WTC) \cite{Sannino:2004qp}. (d) Finally, stable charged
clusters $\bar u_5 \bar u_5 \bar u_5$ of (anti)quarks $\bar u_5$ of
5th family can follow from the approach, unifying spins and charges
\cite{Norma}. Since all these models also predict corresponding +2
charge antiparticles, cosmological scenario should provide mechanism
of their suppression, what can naturally take place in the
asymmetric case, corresponding to excess of -2 charge species,
$X^{--}$. Then their positively charged antiparticles can
effectively annihilate in the early Universe.

If new stable species belong to non-trivial representations of
electroweak SU(2) group, sphaleron transitions at high temperatures
can provide the relationship between baryon asymmetry and excess of
-2 charge stable species, as it was demonstrated in the case of WTC
\cite{KK,KK2,unesco,iwara}.

 After it is formed
in the Standard Big Bang Nucleosynthesis (SBBN), $^4He$ screens the
$X^{--}$ charged particles in composite $(^4He^{++}X^{--})$ {\it
O-helium} ``atoms''
 \cite{I}.
 For different models of $X^{--}$ these "atoms" are also
called ANO-helium \cite{lom,Khlopov:2006dk}, Ole-helium
\cite{Khlopov:2006dk,FKS} or techni-O-helium \cite{KK}. We'll call
them all O-helium ($OHe$) in our further discussion, which follows
the guidelines of \cite{I2}.

In all these forms of O-helium, $X^{--}$ behaves either as lepton or
as specific "heavy quark cluster" with strongly suppressed hadronic
interaction. Therefore O-helium interaction with matter is
determined by nuclear interaction of $He$. These neutral primordial
nuclear interacting objects contribute to the modern dark matter
density and play the role of a nontrivial form of strongly
interacting dark matter \cite{Starkman,McGuire:2001qj}.

Here after a brief review of the qualitative picture of OHe
cosmological evolution \cite{I,FKS,KK,unesco,Khlopov:2008rp} we
concentrate on some open questions in the properties of these dark
atoms and their interaction with matter. This analysis is used in
our second contribution to explain the puzzles of dark matter
searches \cite{DMDA}

\section{Some features of O-helium Universe}

Following \cite{I,lom,Khlopov:2006dk,KK,unesco,iwara,I2} consider
charge asymmetric case, when excess of $X^{--}$ provides effective
suppression of positively charged species.

In the period $100\s \le t \le 300\s$  at $100 \keV\ge T \ge T_o=
I_{o}/27 \approx 60 \keV$, $^4He$ has already been formed in the
SBBN and virtually all free $X^{--}$ are trapped by $^4He$ in
O-helium ``atoms" $(^4He^{++} X^{--})$. Here the O-helium ionization
potential is\footnote{The account for charge distribution in $He$
nucleus leads to smaller value $I_o \approx 1.3 \MeV$
\cite{Pospelov}.} \beq I_{o} = Z_{x}^2 Z_{He}^2 \alpha^2 m_{He}/2
\approx 1.6 \MeV,\label{IO}\eeq where $\alpha$ is the fine structure
constant,$Z_{He}= 2$ and $Z_{x}= 2$ stands for the absolute value of
electric charge of $X^{--}$.  The size of these ``atoms" is
\cite{I,FKS} \beq R_{o} \sim 1/(Z_{x} Z_{He}\alpha m_{He}) \approx 2
\cdot 10^{-13} \cm \label{REHe} \eeq Here and further, if not
specified otherwise, we use the system of units $\hbar=c=k=1$.

Due to nuclear interactions of its helium constituent with nuclei in
the cosmic plasma, the O-helium gas is in thermal equilibrium with
plasma and radiation on the Radiation Dominance (RD) stage, while
the energy and momentum transfer from plasma is effective. The
radiation pressure acting on the plasma is then transferred to
density fluctuations of the O-helium gas and transforms them in
acoustic waves at scales up to the size of the horizon.

At temperature $T < T_{od} \approx 200 S^{2/3}_3\eV$ the energy and
momentum transfer from baryons to O-helium is not effective
\cite{I,KK} because $$n_B \sv (m_p/m_o) t < 1,$$ where $m_o$ is the
mass of the $OHe$ atom and $S_3= m_o/(1 \TeV)$. Here \beq \sigma
\approx \sigma_{o} \sim \pi R_{o}^2 \approx
10^{-25}\cm^2\label{sigOHe}, \eeq and $v = \sqrt{2T/m_p}$ is the
baryon thermal velocity. Then O-helium gas decouples from plasma. It
starts to dominate in the Universe after $t \sim 10^{12}\s$  at $T
\le T_{RM} \approx 1 \eV$ and O-helium ``atoms" play the main
dynamical role in the development of gravitational instability,
triggering the large scale structure formation. The composite nature
of O-helium determines the specifics of the corresponding dark
matter scenario.

At $T > T_{RM}$ the total mass of the $OHe$ gas with density $\rho_d
= (T_{RM}/T) \rho_{tot} $ is equal to
$$M=\frac{4 \pi}{3} \rho_d t^3 = \frac{4 \pi}{3} \frac{T_{RM}}{T} m_{Pl}
(\frac{m_{Pl}}{T})^2$$ within the cosmological horizon $l_h=t$. In
the period of decoupling $T = T_{od}$, this mass  depends strongly
on the O-helium mass $S_3$ and is given by \cite{KK}\beq M_{od} =
\frac{T_{RM}}{T_{od}} m_{Pl} (\frac{m_{Pl}}{T_{od}})^2 \approx 2
\cdot 10^{44} S^{-2}_3 \g = 10^{11} S^{-2}_3 M_{\odot}, \label{MEPm}
\eeq where $M_{\odot}$ is the solar mass. O-helium is formed only at
$T_{o}$ and its total mass within the cosmological horizon in the
period of its creation is $M_{o}=M_{od}(T_{od}/T_{o})^3 = 10^{37}
\g$.

On the RD stage before decoupling, the Jeans length $\lambda_J$ of
the $OHe$ gas was restricted from below by the propagation of sound
waves in plasma with a relativistic equation of state
$p=\epsilon/3$, being of the order of the cosmological horizon and
equal to $\lambda_J = l_h/\sqrt{3} = t/\sqrt{3}.$ After decoupling
at $T = T_{od}$, it falls down to $\lambda_J \sim v_o t,$ where $v_o
= \sqrt{2T_{od}/m_o}.$ Though after decoupling the Jeans mass in the
$OHe$ gas correspondingly falls down
$$M_J \sim v_o^3 M_{od}\sim 3 \cdot 10^{-14}M_{od},$$ one should
expect a strong suppression of fluctuations on scales $M<M_o$, as
well as adiabatic damping of sound waves in the RD plasma for scales
$M_o<M<M_{od}$. It can provide some suppression of small scale
structure in the considered model for all reasonable masses of
O-helium. The significance of this suppression and its effect on the
structure formation needs a special study in detailed numerical
simulations. In any case, it can not be as strong as the free
streaming suppression in ordinary Warm Dark Matter (WDM) scenarios,
but one can expect that qualitatively we deal with Warmer Than Cold
Dark Matter model.

Being decoupled from baryonic matter, the $OHe$ gas does not follow
the formation of baryonic astrophysical objects (stars, planets,
molecular clouds...) and forms dark matter halos of galaxies. It can
be easily seen that O-helium gas is collisionless for its number
density, saturating galactic dark matter. Taking the average density
of baryonic matter one can also find that the Galaxy as a whole is
transparent for O-helium in spite of its nuclear interaction. Only
individual baryonic objects like stars and planets are opaque for
it.

\section{Signatures of O-helium dark matter in the Galaxy}
The composite nature of O-helium dark matter results in a number of
observable effects, which we briefly discuss following
\cite{unesco}.
\subsection{Anomalous component of cosmic rays}
O-helium atoms can be destroyed in astrophysical processes, giving
rise to acceleration of free $X^{--}$ in the Galaxy.

O-helium can be ionized due to nuclear interaction with cosmic rays
\cite{I,I2}. Estimations \cite{I,Mayorov} show that for the number
density of cosmic rays $ n_{CR}=10^{-9}\cm^{-3}$ during the age of
Galaxy a fraction of about $10^{-6}$ of total amount of OHe is
disrupted irreversibly, since the inverse effect of recombination of
free $X^{--}$ is negligible. Near the Solar system it leads to
concentration of free $X^{--}$ $ n_{X}= 3 \cdot 10^{-10}S_3^{-1}
\cm^{-3}.$ After OHe destruction free $X^{--}$ have momentum of
order $p_{X} \cong \sqrt{2 \cdot M_{X} \cdot I_{o}} \cong 2 \GeV
S_3^{1/2}$ and velocity $v/c \cong 2 \cdot 10^{-3} S_3^{-1/2}$ and
due to effect of Solar modulation these particles initially can
hardly reach Earth \cite{KK2,Mayorov}. Their acceleration by Fermi
mechanism or by the collective acceleration forms power spectrum of
$X^{--}$ component at the level of $X/p \sim n_{X}/n_g = 3 \cdot
10^{-10}S_3^{-1},$ where $n_g \sim 1 \cm^{-3}$ is the density of
baryonic matter gas.

At the stage of red supergiant stars have the size $\sim 10^{15}
\cm$ and during the period of this stage$\sim 3 \cdot 10^{15} \s$,
up to $\sim 10^{-9}S_3^{-1}$ of O-helium atoms per nucleon can be
captured \cite{KK2,Mayorov}. In the Supernova explosion these OHe
atoms are disrupted in collisions with particles in the front of
shock wave and acceleration of free $X^{--}$ by regular mechanism
gives the corresponding fraction in cosmic rays. However, this
picture needs detailed analysis, based on the development of OHe
nuclear physics and numerical studies of OHe evolution in the
stellar matter.

If these mechanisms of $X^{--}$ acceleration are effective, the
anomalous low $Z/A$ component of $-2$ charged $X^{--}$ can be
present in cosmic rays at the level $X/p \sim n_{X}/n_g \sim
10^{-9}S_3^{-1},$ and be within the reach for PAMELA and AMS02
cosmic ray experiments.

In the framework of Walking Technicolor model the excess of both
stable $X^{--}$ and $Y^{++}$ is possible \cite{KK2}, the latter
being two-three orders of magnitude smaller, than the former. It
leads to the two-component composite dark matter scenario with the
dominant OHe accompanied by a subdominant WIMP-like component of
$(X^{--}Y^{++})$ bound systems. Technibaryons and technileptons can
be metastable and decays of $X^{--}$ and $Y^{++}$ can provide
explanation for anomalies, observed in high energy cosmic positron
spectrum by PAMELA and in high energy electron spectrum by FERMI and
ATIC.

\subsection{Positron annihilation and gamma lines in galactic
bulge} Inelastic interaction of O-helium with the matter in the
interstellar space and its de-excitation can give rise to radiation
in the range from few keV to few  MeV. In the galactic bulge with
radius $r_b \sim 1 \kpc$ the number density of O-helium can reach
the value $n_o\approx 3 \cdot 10^{-3}/S_3 \cm^{-3}$ and the
collision rate of O-helium in this central region was estimated in
\cite{I2}: $dN/dt=n_o^2 \sigma v_h 4 \pi r_b^3 /3 \approx 3 \cdot
10^{42}S_3^{-2} \s^{-1}$. At the velocity of $v_h \sim 3 \cdot 10^7
\cm/\s$ energy transfer in such collisions is $\Delta E \sim 1 \MeV
S_3$. These collisions can lead to excitation of O-helium. If 2S
level is excited, pair production dominates over two-photon channel
in the de-excitation by $E0$ transition and positron production with
the rate $3 \cdot 10^{42}S_3^{-2} \s^{-1}$ is not accompanied by
strong gamma signal. According to \cite{Finkbeiner:2007kk} this rate
of positron production for $S_3 \sim 1$ is sufficient to explain the
excess in positron annihilation line from bulge, measured by
INTEGRAL (see \cite{integral} for review and references). If $OHe$
levels with nonzero orbital momentum are excited, gamma lines should
be observed from transitions ($ n>m$) $E_{nm}= 1.598 \MeV (1/m^2
-1/n^2)$ (or from the similar transitions corresponding to the case
$I_o = 1.287 \MeV $) at the level $3 \cdot 10^{-4}S_3^{-2}(\cm^2 \s
\MeV ster)^{-1}$.

It should be noted that the nuclear cross section of the O-helium
interaction with matter escapes the severe constraints
\cite{McGuire:2001qj} on strongly interacting dark matter particles
(SIMPs) \cite{Starkman,McGuire:2001qj} imposed by the XQC experiment
\cite{XQC}. Therefore, a special strategy of direct O-helium  search
is needed, as it was proposed in \cite{Belotsky:2006fa}.

\section{O-helium interaction with nuclei}
The evident consequence of the O-helium dark matter is its
inevitable presence in the terrestrial matter, which appears opaque
to O-helium and stores all its in-falling flux. After they fall down
terrestrial surface, the in-falling $OHe$ particles are effectively
slowed down due to elastic collisions with matter.In underground
detectors, $OHe$ ``atoms'' are slowed down to thermal energies and
give rise to energy transfer $\sim 2.5 \cdot 10^{-4} \eV A/S_3$, far
below the threshold for direct dark matter detection. It makes this
form of dark matter insensitive to the severe CDMS constraints
\cite{Akerib:2005kh}. However, $OHe$ induced processes in the matter
of underground detectors can result in observable effects. These
effects, considered in a separate contribution \cite{DMDA}, strongly
depend on the details of the OHe interaction with nuclei, which we
consider here.

\subsection{Structure of $X^{--}$ atoms with nuclei}
The properties of OHe interaction with matter are determined first
of all by the structure of OHe atom that follows from the general
analysis of the bound states of $X^{--}$ with nuclei.

Consider a simple model \cite{Pospelov}, in which the nucleus is
regarded as a sphere with uniform charge density and in which the
mass of the $X^{--}$ is assumed to be much larger than that of the
nucleus. Spin dependence is also not taken into account so that both
the particle and nucleus are considered as scalars. Then the
Hamiltonian is given by
\begin{equation}
    H=\frac{p^2}{2 A m_p} - \frac{Z Z_x \alpha}{2 R} + \frac{Z Z_x \alpha}{2 R} \cdot (\frac{r}{R})^2,
\end{equation}
for short distances $r<R$ and
\begin{equation}
    H=\frac{p^2}{2 A m_p} - \frac{Z Z_x \alpha}{R},
\end{equation}
for long distances $r>R$, where $\alpha$ is the fine structure
constant, $R = d_o A^{1/3} \sim 1.2 A^{1/3} /(200 MeV)$ is the
nuclear radius, $Z$ is the electric charge of nucleus and $Z_x=2$ is
the electric charge of negatively charged particle $X^{--}$. Since
$A m_p \ll M_X$ the reduced mass is $1/m= 1/(A m_p) + 1/M_X \approx
1/(A m_p)$.

For small nuclei the Coulomb binding energy is like in hydrogen atom
and is given by
\begin{equation}
    E_b=\frac{1}{2} Z^2 Z_x^2 \alpha^2 A m_p.
\end{equation}

For large nuclei $X^{--}$ is inside nuclear radius and the harmonic
oscillator approximation is valid for the estimation of the binding
energy
\begin{equation}
    E_b=\frac{3}{2}(\frac{Z Z_x \alpha}{R}-\frac{1}{R}(\frac{Z Z_x \alpha}{A m_p R})^{1/2}).
\label{potosc}
\end{equation}

For the intermediate regions between these two cases with the use of
trial function of the form $\psi \sim e^{- \gamma r/R}$ variational
treatment of the problem \cite{Pospelov} gives
\begin{equation}
    E_b=\frac{1}{A m_p R^2} F(Z Z_x \alpha A m_p R ),
\end{equation}
where the function $F(a)$ has limits
\begin{equation}
    F(a \rightarrow 0) \rightarrow \frac{1}{2}a^2  - \frac{2}{5} a^4
\end{equation}
and
\begin{equation}
    F(a \rightarrow \infty) \rightarrow \frac{3}{2}a  - (3a)^{1/2},
\end{equation}
where $a = Z Z_x \alpha A m_p R$. For $0 < a < 1$ the Coulomb model
gives a good approximation, while at $2 < a < \infty$ the harmonic
oscillator approximation is appropriate.

In the case of OHe $a = Z Z_x \alpha A m_p R \le 1$, what proves its
Bohr-atom-like structure, assumed in our earlier papers
\cite{I,lom,Khlopov:2006dk,KK,unesco,iwara,I2}. However, the size of
He, rotating around $X^{--}$ in this Bohr atom, turns out to be of
the order and even a bit larger than the radius $r_o$ of its Bohr
orbit, and the corresponding correction to the binding energy due to
non-point-like charge distribution in He is significant.

Bohr atom like structure of OHe seems to provide a possibility to
use the results of atomic physics for description of OHe interaction
with matter. However, the situation is much more complicated. OHe
atom is similar to the hydrogen, in which electron is hundreds times
heavier, than proton, so that it is proton shell that surrounds
"electron nucleus". Nuclei that interact with such "hydrogen" would
interact first with strongly interacting "protonic" shell and such
interaction can hardly be treated in the framework of perturbation
theory. Moreover in the description of OHe interaction the account
for the finite size of He, which is even larger than the radius of
Bohr orbit, is important. One should consider, therefore, the
analysis, presented below, as only a first step approaching true
nuclear physics of OHe.

\subsection{Potential of O-helium interaction with nuclei}

Our explanation \cite{unesco,iwara,Bled09} of the results of
DAMA/NaI \cite{Bernabei:2003za} and DAMA/LIBRA
\cite{Bernabei:2008yi} experiments is based on the idea that OHe,
slowed down in the matter of detector, can form a few keV bound
state with nucleus, in which OHe is situated \textbf{beyond} the
nucleus. Therefore the positive result of these experiments is
explained by reaction
\begin{equation}
A+(^4He^{++}X^{--}) \rightarrow [A(^4He^{++}X^{--})]+\gamma
\label{HeEAZ}
\end{equation}
with nuclei in DAMA detector.

In our earlier studies \cite{unesco,iwara,Bled09} the conditions
were found, under which both sodium and iodine nuclei have a few keV
bound states with OHe, explaining the results of DAMA experiments by
OHe radiative capture to these levels. Here we extend the set of our
solutions by the case, when the results of DAMA experiment can be
explained by radiative OHe capture by sodium only and there are no
such bound states with iodine and Tl.

Schrodinger equation for OHe-nucleus system is reduced (taking apart
the equation for the center of mass) to the equation of relative
motion for the reduced mass
\begin{equation}
            m=\frac{Am_p m_o}{Am_p+m_o},
            \label{m}
 \end{equation}
where $m_p$ is the mass of proton and $m_o\approx M_X+4m_p$ is the
mass of OHe. Since $m_o \approx M_X \gg A m_p$, center of mass of
Ohe-nucleus system approximately coincides with the position of
$X^{--}$.

In the case of orbital momentum \emph{l}=0 the wave functions depend
only on \emph{r}.

The approach of \cite{unesco,iwara,Bled09} assumes the following
picture: at the distances larger, than its size, OHe is neutral,
being only the source of a Coulomb field of $X^{--}$ screened by
$He$ shell
\begin{equation}
U_c= \frac{Z_{X} Z \alpha  \cdot F_X(r)}{r}, \label{epotem}
\end{equation}
where $Z_{X}=-2$ is the charge of $X^{--}$, $Z$ is charge of
nucleus, $F_X(r)=(1+r/r_o) exp(-2r/r_o)$ is the screening factor of
Coulomb potential (see e.g.\cite{LL3}) of $X^{--}$ and $r_o$ is the
size of OHe. Owing to the negative sign of $Z_{X}=-2$, this
potential provides attraction of nucleus to OHe.

Then helium shell of OHe starts to feel Yukawa exponential tail of
attraction of nucleus to $He$ due to scalar-isoscalar nuclear
potential. It should be noted that scalar-isoscalar nature of He
nucleus excludes its nuclear interaction due to $\pi$ or $\rho$
meson exchange, so that the main role in its nuclear interaction
outside the nucleus plays $\sigma$ meson exchange, on which nuclear
physics data are not very definite. The nuclear potential depends on
the relative distance between He and nucleus and we take it in the
form
\begin{equation}
U_n=-\frac{A_{He} A g^2 exp{(-\mu
|\vec{r}-\vec{\rho}|)}}{|\vec{r}-\vec{\rho}|}. \label{epotnuc}
\end{equation}
Here $\vec{r}$ is radius vector to nucleus, $\vec{\rho}$ is the
radius vector to He in OHe, $A_{He}=4$ is atomic weight of helium,
$A$ is atomic weight of nucleus, $\mu$ and $g^2$ are the mass and
coupling of $\sigma$ meson - mediator of nuclear attraction.

Strictly speaking, starting from this point we should deal with a
three-body problem for the system of He, nucleus and $X^{--}$ and
the correct quantum mechanical description should be based on the
cylindrical and not spherical symmetry. In the present work we use
the approximation of spherical symmetry and take into account
nuclear attraction beyond the nucleus in a two different ways: 1)
nuclear attraction doesn't influence the structure of OHe, so that
the Yukawa potential (\ref{epotnuc}) is averaged over $|\vec{\rho}|$
for spherically symmetric wave function of He shell in OHe; 2)
nuclear attraction changes the structure of OHe so that He takes the
position $|\vec{\rho}|=r_o$, which is most close to the nucleus. Due
to strong attraction of He by the nucleus the second case (which is
labeled "b" in successive numerical calculations) seems more
probable. In the lack of the exact solution of the problem we
present both the results, corresponding to the first case (which are
labeled "m" in successive numerical calculations), and to the second
case (which is labeled "b") in order to demonstrate high sensitivity
of the numerical results to choice of parameters.

In the both cases nuclear attraction results in the polarization of
OHe and the mutual attraction of nucleus and OHe is changed by
Coulomb repulsion of $He$ shell. Taking into account Coulomb
attraction of nucleus by $X^{--}$ one obtains dipole Coulomb barrier
of the form
\begin{equation}
U_d=\frac{Z_{He} Z \alpha r_o}{r^2}. \label{epotdip}
\end{equation}

When helium is completely merged with the nucleus the interaction is
reduced to the oscillatory potential (\ref{potosc}) of $X^{--}$ with
homogeneously charged merged nucleus with the charge $Z+2$, given by
\begin{equation}
    E_m=\frac{3}{2}(\frac{(Z+2) Z_x \alpha}{R}-\frac{1}{R}(\frac{(Z+2) Z_x \alpha}{(A+4) m_p R})^{1/2}).
\label{potosc}
\end{equation}

\begin{figure}
    \begin{center}
        \includegraphics[width=4in]{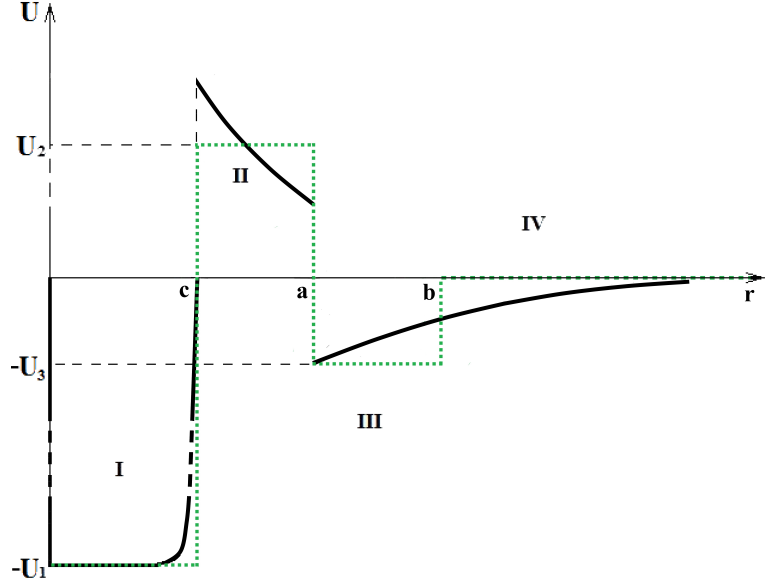}\\
        \caption{The approximation of rectangular well for potential of OHe-nucleus system.}\label{pic1}
    \end{center}
\end{figure}

To simplify the solution of Schrodinger equation we approximate the
potentials (\ref{epotem})-(\ref{potosc}) by a rectangular potential
that consists of a potential well with the depth $U_1$ at $r<c=R$,
where $R$ is the radius of nucleus, of a rectangular dipole Coulomb
potential barrier $U_2$ at $R \le r<a=R+r_o+r_{he}$, where $r_{he}$
is radius of helium nucleus, and of the outer potential well $U_3$,
formed by the Yukawa nuclear interaction (\ref{epotnuc}) and
residual Coulomb interaction (\ref{epotem}). The values of $U_1$ and
$U_2$ were obtained by the averaging of the (\ref{potosc}) and
(\ref{epotdip}) in the corresponding regions, while $U_3$ was equal
to the value of the nuclear potential (\ref{epotnuc}) at $r=a$ and
the width of this outer rectangular well (position of the point b)
was obtained by the integral of the sum of potentials
(\ref{epotnuc}) and (\ref{epotem}) from $a$ to $\infty$.
 It leads to the approximate potential, presented on Fig. \ref{pic1}.

Solutions of Schrodinger equation for each of the four regions,
indicated on Fig. \ref{pic1}, are given in textbooks (see
e.g.\cite{LL3}) and their sewing determines the condition, under
which a low-energy  OHe-nucleus bound state appears in the region
III.
\subsection{Low energy bound state of O-helium with nuclei}
\begin{figure}
    \begin{center}
        \includegraphics[width=6in]{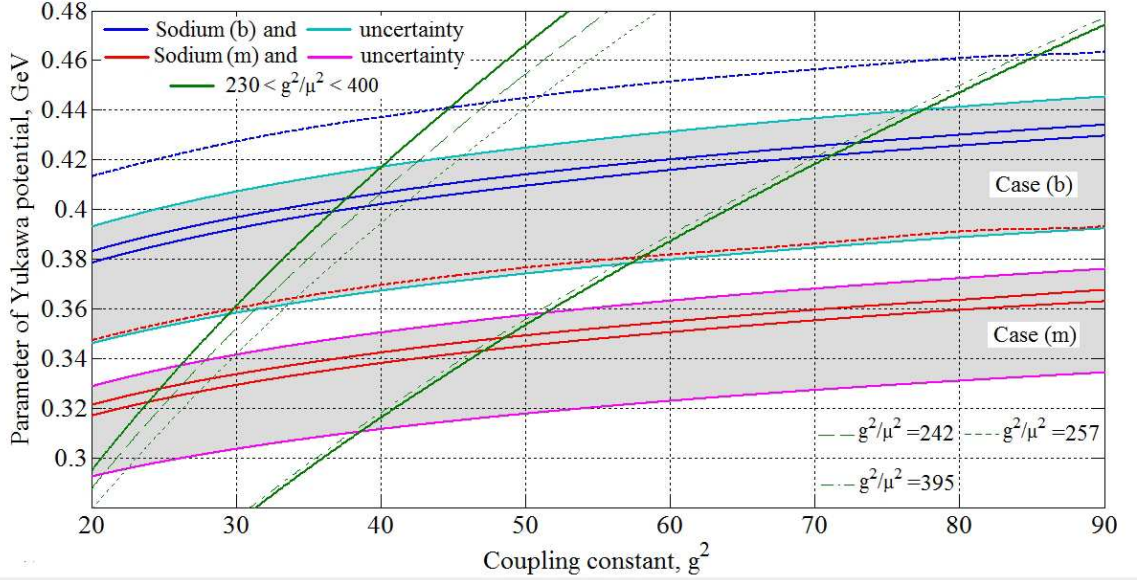}\\
        \caption{
        The region of parameters $\mu$ and $g^2$, for which Na-OHe system has a level in the interval 4 keV.
        Two lines determine at $d_o=1.2/(200 \MeV)$ the region of parameters, at which the bound
        system of this element with OHe has a 4 keV level.
        In the region between the two strips the energy of level is below 4 keV.
        There are also indicated the range of $g^2/\mu^2$ (dashed lines) as well as their preferred values
        (thin lines) determined in \cite{nuclear} from parametrization of the relativistic ($\sigma-\omega$) model for nuclear matter.
        The uncertainty in the determination of parameter $1.15/(200 \MeV)<d_o <1.3/(200 \MeV)$ results in the
        uncertainty of $\mu$ and $g^2$ shown by the shaded regions surrounding the lines.
        The case of nuclear Yukawa potential $U_{3m}$, averaged over the orbit of He in OHe, corresponds to the
        lower lines and shaded region, while the upper lines and shaded region around them illustrate
        the case of nuclear Yukawa potential $U_{3b}$ with the position
        of He most close to the nucleus at $\rho=r_o$.}\label{Na}
    \end{center}
\end{figure}

The energy of this bound state and its existence strongly depend on
the parameters $\mu$ and $g^2$ of nuclear potential (\ref{epotnuc}).
On the Fig. \ref{Na} the regions of these parameters, giving 4 keV
energy level in OHe bound state with sodium are presented. Radiative
capture to this level can explain results of DAMA/NaI and DAMA/LIBRA
experiments with the account for their energy resolution
\cite{DAMAlibra}. The lower shaded region on Fig. \ref{Na}
corresponds to the case of nuclear Yukawa potential $U_{3m}$,
averaged over the orbit of He in OHe, while the upper region
corresponds to the case of nuclear Yukawa potential $U_{3b}$ with
the position of He most close to the nucleus at $\rho=r_o$.The
result is also sensitive to the precise value of $d_o$, which
determines the size of nuclei $R=d_o A^{1/3}$. The two narrow strips
in each region correspond to the experimentally most probable value
$d_o=1.2/(200 \MeV)$. In these calculations the mass of OHe was
taken equal to $m_o=1 TeV$, however the results weakly depend on the
value of $m_o>1 TeV$.

\subsection{Energy levels in other nuclei}
\begin{figure}
\begin{center}
        \includegraphics[width=6in]{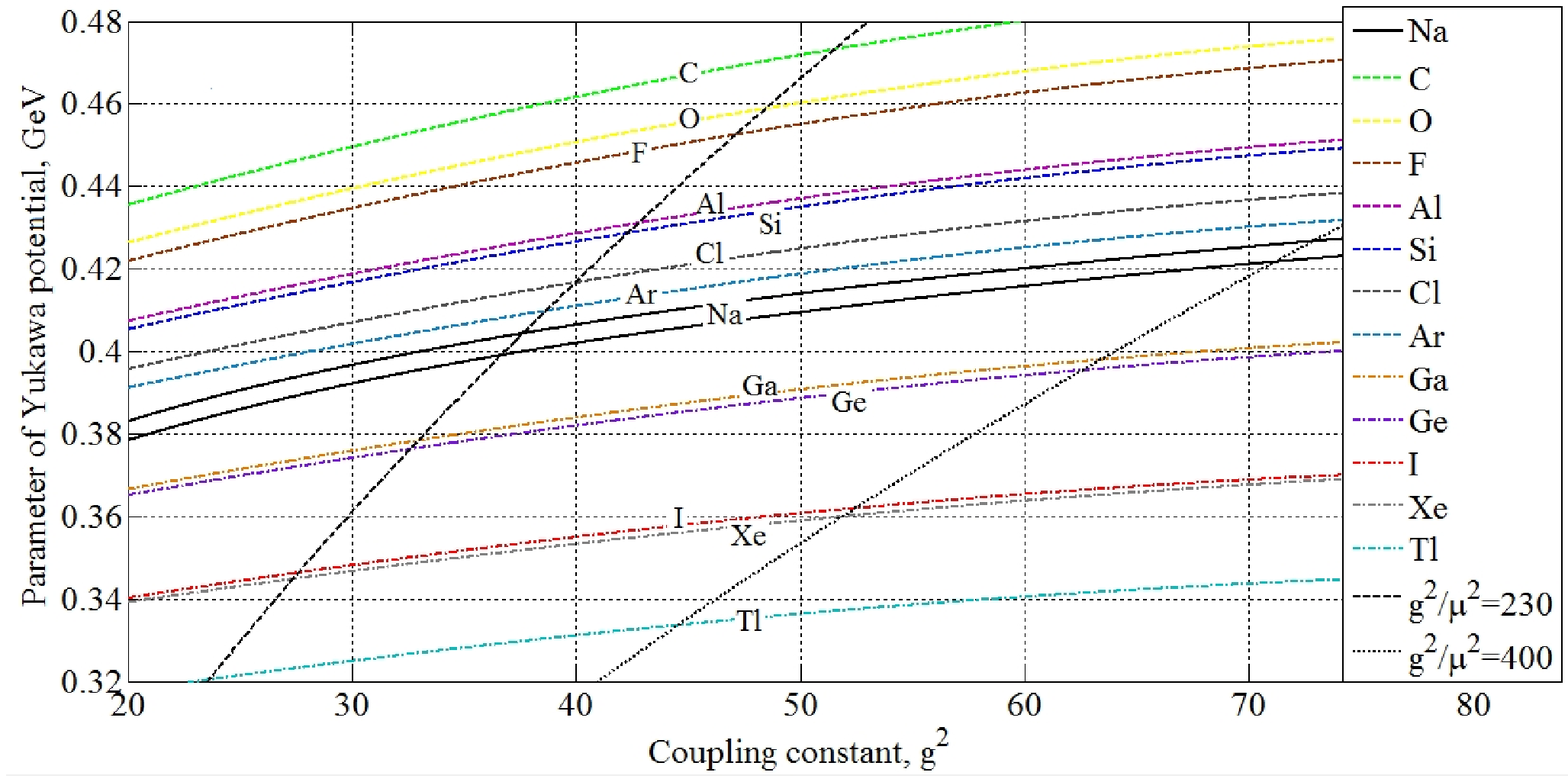}\\
        \caption{Existence of low energy bound states in OHe-nucleus system
        in the case b for nuclear Yukawa potential $U_{3b}$ with the position of
        He most close to the nucleus at $\rho=r_o$.
        The lines, corresponding to different nuclei, show the upper limit
        for nuclear physics parameters $\mu$ and $g^2$, at which these bound states
        are possible. The choice of parameters corresponding to 4 keV OHe-Na
        bound state, excludes region below Na line.}\label{Exb}
    \end{center}
\end{figure}

\begin{figure}
\begin{center}
        \includegraphics[width=6in]{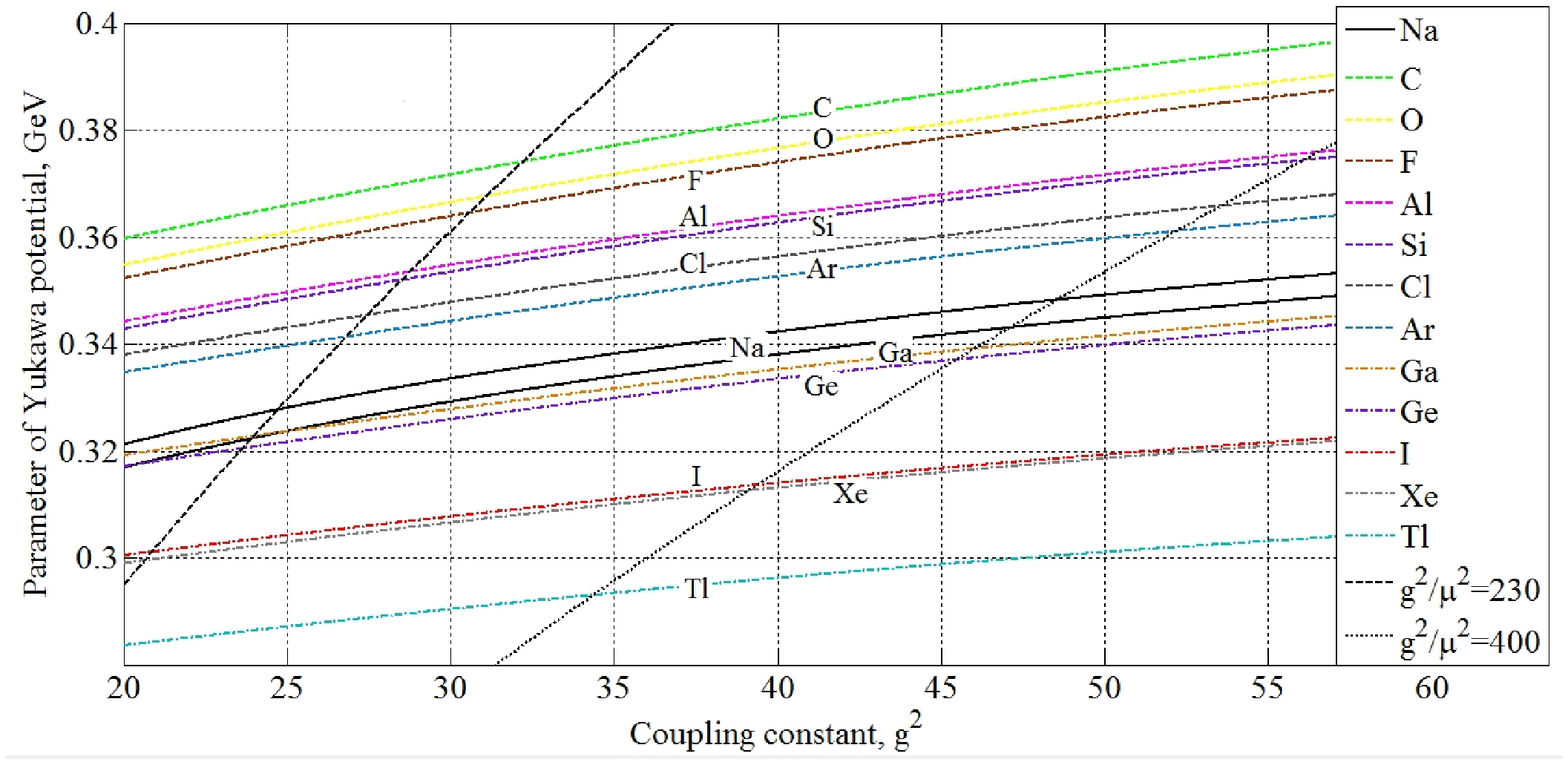}\\
        \caption{Existence of low energy bound states in OHe-nucleus system
        in the case m for nuclear Yukawa potential $U_{3m}$, averaged over the orbit of He in OHe.
        The lines, corresponding to different nuclei, show the upper limit
        for nuclear physics parameters $\mu$ and $g^2$, at which these bound states
        are possible. The choice of parameters corresponding to 4 keV OHe-Na
        bound state, excludes region below Na line.}\label{Exm}
    \end{center}
\end{figure}

The important qualitative feature of the presented solution is the
restricted range of intermediate nuclei, in which the OHe-nucleus
state beyond nuclei is possible. For the chosen range of nuclear
parameters, reproducing the results of DAMA/NaI and DAMA/LIBRA, we
can calculate the binding energy of OHe-nucleus states in nuclei,
corresponding to chemical composition of set-ups in other
experiments. It turns out that  there are no such states for light
and heavy nuclei. In the case of nuclear Yukawa potential $U_{3b}$,
corresponding to the position of He most close to the nucleus at
$\rho=r_o$, the range of nuclei with bound states with OHe
corresponds to the part of periodic table between B and Ti. This
result is stable independent on the used scheme of numerical
calculations. The upper limits on the nuclear parameters $\mu$ and
$g^2$, at which there exists OHe-nucleus bound state are presented
for this case on Fig.\ref{Exb}. In the case of potential $U_{3m}$,
averaged over the orbit of He in OHe, there are no OHe bound states
with nuclei, lighter than Be and heavier than Ti. However, the
results are very sensitive to the numerical factors of calculations
and the existence of OHe-Ge and OHe-Ga bound states at a narrow
window of parameters $\mu$ and $g^2$ turns to be strongly dependent
on these factors so that change in numbers smaller than 1\% can give
qualitatively  different result for Ge and Ga. The results for the
case (m) are shown on Fig.\ref{Exm}.
Both for the cases (b) and (m) there is a stable conclusion that
there are no OHe-nucleus bound states with Xe, I and Tl.

\begin{figure}
    \begin{center}
        \includegraphics[width=6in]{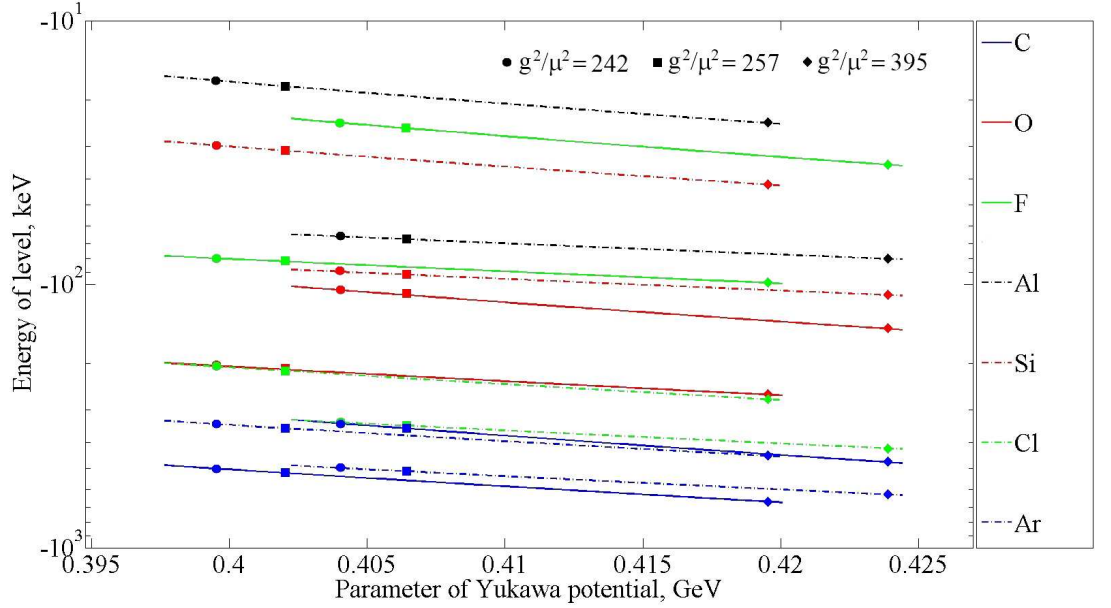}\\
        \caption{Energy levels in OHe bound system with carbon,
        oxygen, fluorine, argon, silicon, aluminium and chlorine for the case of the nuclear
Yukawa potential $U_{3b}$.
The predictions are given for the range
of $g^2/\mu^2$ determined in \cite{nuclear} from parametrization of
the relativistic ($\sigma-\omega$) model for nuclear matter. The
preferred values of $g^2/\mu^2$ are indicated by the corresponding
marks (squares or circles)}\label{elementsb}
    \end{center}
\end{figure}

\begin{figure}
\begin{center}
        \includegraphics[width=6in]{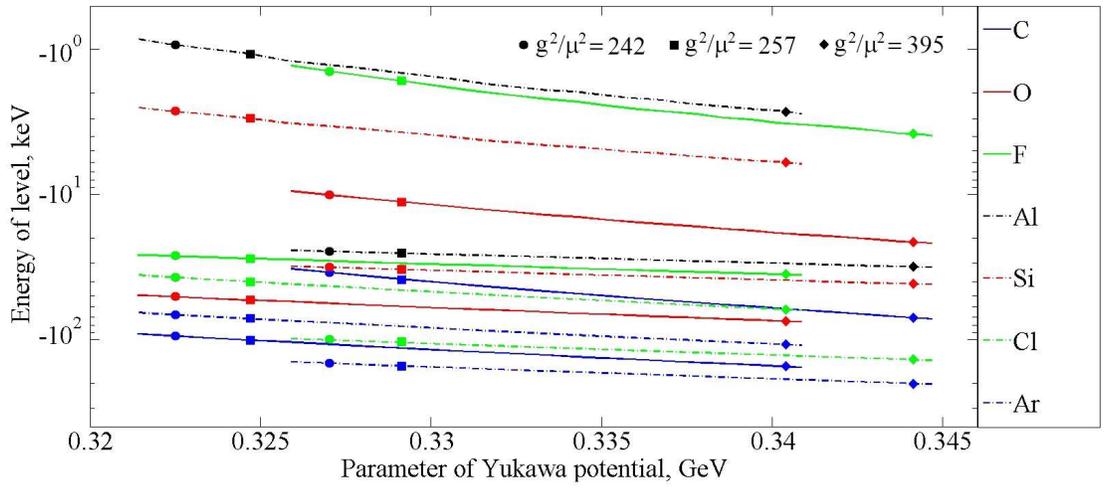}\\
        \caption{Energy levels in OHe bound system with carbon,
        oxygen, fluorine, argon, silicon, aluminium and chlorine for the case of the nuclear
Yukawa potential $U_{3m}$.
The predictions are given for the range
of $g^2/\mu^2$ determined in \cite{nuclear} from parametrization of
the relativistic ($\sigma-\omega$) model for nuclear matter. The
preferred values of $g^2/\mu^2$ are indicated by the corresponding
marks (squares or circles)}\label{elementsm}
    \end{center}
\end{figure}

For the experimentally preferred value $d_o=1.2/(200 \MeV)$ the
results of calculation of the binding energy of OHe-nucleus systems
for carbon, oxygen, fluorine, argon, silicon, aluminium and chlorine
are presented on Fig. \ref{elementsb} for the case of the nuclear
Yukawa potential $U_{3b}$ and on Fig. \ref{elementsm} for the case
of the potential $U_{3m}$. The difference in these results
demonstrates their high sensitivity to the choice of parameters.

\section{Conclusions}


To conclude, the results of dark matter search in experiments
DAMA/NaI and DAMA/LIBRA can be explained in the framework of
composite dark matter scenario without contradiction with negative
results of other groups. This scenario can be realized in different
frameworks, in particular in Minimal Walking Technicolor model or in
the approach unifying spin and charges and contains distinct
features, by which the present explanation can be distinguished from
other recent approaches to this problem \cite{Edward} (see also
review and more references in \cite{Gelmini}).

Our explanation is based on the mechanism of low energy binding of
OHe with nuclei. We have found that within the uncertainty of
nuclear physics parameters there exists their range at which OHe
binding energy with sodium is equal to 4 keV and there is no such
binding with iodine and thallium.


With the account for high sensitivity of our results to the values
of uncertain nuclear parameters and for the approximations, made in
our calculations, the presented results can be considered only as an
illustration of the possibility to explain effects in underground
detectors by OHe binding with intermediate nuclei. However, even at
the present level of our studies we can make a conclusion that
effects of such binding should strongly differ in detectors with the
content, different from NaI, and can be absent in detectors with
very light (e.g. $^3He$) and heavy nuclei (like xenon and probably
germanium). Therefore test of results of DAMA/NaI and DAMA/LIBRA
experiments by other experimental groups can become a very
nontrivial task.



\section {Acknowledgments}


We would like to thank Norma Mankoc-Borstnik and all the
participants of Bled Workshop for stimulating discussions.





\end{document}